\documentclass[aps,twocolumn,showpacs,prl,superscriptaddress,floatfix]{revtex4}
\usepackage{graphicx}
\usepackage{epsf}
\usepackage{wrapfig}
\usepackage{epsfig}
\usepackage{sublabel}
\usepackage{epsfig}
\usepackage{verbatim}
\usepackage{amsmath}

\def\b0{{\mbox{\boldmath$0$}}}

\usepackage{graphicx}
\usepackage{epsf}
\usepackage{wrapfig}
\usepackage{epsfig}
\def\Vec#1{\mbox{\boldmath $#1$}}

\def\beq{\begin{equation}}
\def\eeq{\end{equation}}

\def\beqy{\begin{eqnarray}}
\def\eeqy{\end{eqnarray}}

\def \b #1{ {\bf #1}}
\newcommand{\be}{\begin{eqnarray}}
\newcommand{\ee}{\end{eqnarray}}

\def \b #1{ {\bf #1}}

\def \b #1{ {\bf #1}}
     \font\tenbifull=cmmib10 scaled 1200 
     \font\tenbimed=cmmib9
     \font\tenbismall=cmmib7
\mathchardef\bbkappa="7114
\mathchardef\bbrho="711A
\mathchardef\bbsigma="711B
\mathchardef\bbtau="711C
\mathchardef\bbvarrho="7125
\mathchardef\bbvarsigma="7126
\mathchardef\bbxi="7118

\begin{document}
\vskip 2mm \date{\today}\vskip 2mm
\title{Universality of nucleon-nucleon short-range correlations: two-nucleon momentum distributions in few-body systems.
}
\author{M. Alvioli}
\affiliation{ECT$^\star$, European Center for Theoretical Studies in Nuclear Physics
and Related Areas,\\ Strada delle Tabarelle 286, I-38123 Villazzano (TN) Italy}
\author{C. Ciofi degli Atti}
   \author{L. P. Kaptari}
   \altaffiliation{On leave from the Bogolubov Lab. Theor. Phys., JINR, 141980 Dubna, Russia,
   through the program Rientro dei Cervelli of the Italian Ministry
of University and Research}
\author{C. B. Mezzetti}
\affiliation{Department of Physics, University of Perugia and\\
  Istituto Nazionale di Fisica Nucleare, Sezione di Perugia\\
  Via A. Pascoli, I-06123, Italy}
\author{H. Morita}\affiliation{Sapporo Gakuin University, Bunkyo-dai 11, Ebetsu 069-8555,
  Hokkaido, Japan}
\author{S. Scopetta}
\affiliation{Department of Physics, University of Perugia and\\
  Istituto Nazionale di Fisica Nucleare, Sezione di Perugia\\
  Via A. Pascoli, I-06123, Italy}

\vskip 2mm
\begin{abstract}
Using realistic  wave functions, the proton-neutron and
proton-proton momentum distributions in $^3He$ and $^4He$ are
calculated as a function of the relative, $k_{rel}$, and center of
mass, $K_{CM}$, momenta, and the angle between them. For large
values of ${k}_{rel}\gtrsim 2\,\,fm^{-1}$ and small values of  ${
K}_{CM} \lesssim 1.0\,\,fm^{-1}$,  both distributions are angle
independent and decrease with increasing $K_{CM}$, with  the $pn$
distribution factorizing into the deuteron momentum distribution
times a rapidly decreasing function of $K_{CM}$, in agreement with
the two-nucleon (2N) short range correlation (SRC) picture.  When
$K_{CM}$ and $k_{rel}$ are both large, the distributions exhibit a
strong angle dependence, which is evidence of three-nucleon (3N)
SRC. The predicted center-of-mass and angular dependence of 2N and
3N SRC should be observable in two-nucleon knock-out processes
$A(e,e'pN)X$.
\end{abstract}

 \maketitle
Realistic many-body calculations (see e.g.
\cite{Pieper:2001mp}-\cite{Alvioli:2007zz}) show that a mean field
approach, though describing very successfully  many properties of
nuclei, breaks down when the relative distance  $r\equiv |{\Vec
r}_1 - {\Vec r}_2|$ between two generic nucleons "1" and "2"  is
of the order of $r \lesssim 1.3 -1.5 \,\,fm$. In this region
nucleon-nucleon (NN) motion exhibits SRC, arising from the
interplay between  the   short range repulsion and the
intermediate range tensor attraction of the NN potential. As a
result of such an interplay,  the two-nucleon density distribution
strongly deviates from the mean field distribution, in that
whereas the latter has
 a maximum value at zero  separation, the former
 almost vanishes
at $r=0$,  increases sharply with increasing separation,
overshoots at $r \gtrsim 1.3-1.5 \,\,fm$  the mean field density,
and coincides with it at larger separations.
 The detailed structure of SRC depends upon the spin-isospin
state of the NN pair, as well as upon the value of the pair
center-of-mass  (CM) coordinate   ${\Vec R}=({\Vec r}_1 + {\Vec
r}_2)/2$. The study of
  SRC represents one of the main challenges of
nowadays  nuclear physics,  since the detailed theoretical and
experimental knowledge  of the short range structure of nuclei,
could provide decisive answers to longstanding fundamental
questions, such as the formation and structure of cold dense
nuclear matter,  the origin of the EMC effect, the role of
quark-gluon degrees of freedom in nuclei (see e.g.
\cite{Frankfurt:2008zv}). SRC  generate high momentum components,
which are lacking in a mean field approach,  and give rise to
peculiar configurations of the nuclear wave function in momentum
space \cite{Frankfurt:1988nt}. In particular, if nucleons "1" and
"2" become strongly correlated  at short distances, the local
configuration (in the nucleus CM frame) characterized by ${\Vec
k}_2 \simeq -{\Vec k}_1$, ${\Vec K}_{A-2}= \sum_{i=3}^A \,{\Vec
k}_i \simeq 0$,  dominates over the average mean field
configuration $\sum_{i=2}^A \,{\Vec k}_i \simeq -{\Vec k}_{1}$,
which is the configuration  when the high momentum nucleon is
balanced by all of the remaining $A-1$ nucleons. Thus,  if a
correlated nucleon with momentum ${\Vec k}_1$ acquires a momentum
${\Vec q}$ from an external probe and it is removed from the
nucleus and detected with momentum ${\Vec p}= {\Vec k}_1+{\Vec
q}$, the partner nucleon should be emitted with  momentum ${\Vec
k}_2 \simeq {-\Vec k}_1 ={\Vec q}-{\Vec p}\equiv{\Vec p}_{miss}$.
Such a qualitative  picture is strictly valid only if the CM
momentum of the correlated pair was zero before nucleon removal
and, moreover, if the two correlated nucleons leave the nucleus
without interacting between themselves and with the nucleus
$(A-2)$. Nonetheless, recent experimental data on  nucleon
knockout from carbon using protons \cite{Tang:2002ww},
\cite{Piasetzky:2006ai} and electron \cite{Subedi:2008zz}
projectiles, have shown that the removal of a proton from the
nucleus in the range of $0.275 \lesssim |{\Vec p}_{miss}|\lesssim
0.550 \,\,GeV/c$ is almost always accompanied by the emission of a
neutron carrying momentum equal to ${\Vec p}_{miss}$, with a
momentum spread in agreement with a Gaussian motion of the CM of
the correlated pair in the nucleus, as predicted long ago in Ref.
\cite{CiofidegliAtti:1995qe}. Whereas experiments demonstrating
the presence of  SRC in nuclei and their basic mechanism  have
eventually been  performed, detailed information  through the
periodic Table of their isospin, angular and CM dependencies is
still to come. A partial relevant progress has however already
been done  by demonstrating
\cite{Schiavilla:2006xx,Alvioli:2007zz}, in qualitative agreement
with the experimental data on $^{12}$C, that  the strong
 correlations induced by the tensor force lead
to large differences in the $pp$ and $pn$ distributions at
moderate values of the relative momentum of the pair. Such a
result has been confirmed in a recent thorough analysis
\cite{Feldmeier} of the relative (integrated over the pair CM
variables ${\bf R}$ and  ${\bf K}_{CM}={\bf k}_1+{\bf k}_2$)
two-body densities and momentum distributions, and their detailed
dependence upon the spin-isospin states. As for the angular and CM
dependencies of SRC, in Ref.
\cite{Schiavilla:2006xx,Alvioli:2007zz}  the focus was on the
two-body momentum distributions integrated either over the CM or
the relative momenta, whereas   in Ref. \cite{Wiringa:2008dn} the
CM dependence of the relative momentum distributions of $^3He$ and
$^4He$ has been investigated in a particular angular
configuration, namely when ${\Vec K}_{CM}$ and ${\Vec
k}_{rel}\equiv ({\bf k}_1-{\bf k}_2)/2$ are parallel. In this
Letter the results of calculations for arbitrary mutual
orientations of ${\Vec K}_{CM}$ and ${\Vec k}_{rel}$, are
presented and several universal features of SRC will be
demonstrated.  Our calculations were performed with nuclear wave
functions \cite{Kievsky:1992um,akaishi} obtained from the solution
of the Schr\"odinger equation containing realistic NN
interactions, namely the $AV18$ \cite{Wiringa:1994wb} and $AV8'$
\cite{Pudliner:1997ck} interactions. We will compare our results
with a preliminary analysis of data on $^3$He from the CEBAF large
acceptance spectrometer (CLAS) collaboration at
JLab~\cite{Weinstein08}. The summed over spin and isospin two body
momentum distributions of a nucleon-nucleon pair is defined as
follows
\begin{widetext}
\beqy
n^{NN}(\Vec{k}_{1},\Vec{k}_{2})=
\,\frac{1}{(2\pi)^6}\int
d\Vec{r_1}\,d\Vec{r_2}\,d\Vec{r_1}^\prime\,d\Vec{r_2}^\prime\,
e^{i\,\Vec{k}_{1}\cdot\left(\Vec{r}_1-\Vec{r}_1^\prime\right)}\,
e^{i\,\Vec{k}_{2}\cdot\left(\Vec{r}_2-\Vec{r}_2^\prime\right)}\,
\rho^{(2)}_{NN}(\Vec{r}_1,\Vec{r}_2;\Vec{r}_1^\prime,\Vec{r}_2^\prime)\,
\label{2bmomdis} \eeqy
\end{widetext}
with
\begin{widetext}
\beqy
\rho^{(2)}_{NN}(\Vec{r}_1,\Vec{r}_2;\Vec{r}_1^\prime,\Vec{r}_2^\prime)=\frac{1}{2J+1}\sum_M
\int \psi_{JM}^{*}(\Vec r_{1},\Vec r_{2},\Vec
r_{3}...,\Vec{r}_A)\, \psi_{JM}(\Vec r_{1}^{\prime},\Vec
r_{2}^{\prime},\Vec r_{3},...,\Vec{r}_A)\,\delta\left (
 \sum^A_{i=1}\Vec{r}_i\right)\, \prod\displaylimits_{i=3}^A
d\Vec{r}_i \label{ground}
\eeqy
\end{widetext}
being the two-body non diagonal density matrix. In
Eq.~(\ref{ground}) $ \psi_{JM}(\Vec r_{1},\Vec r_{2},\Vec
r_{3}...,\Vec{r}_A)$  is the  wave function of the nucleus in the
ground state with the total angular momentum $J$ and its
projection $M$. The two-nucleon momentum distribution can then be
defined as follows
\begin{widetext}
\beqy
n^{NN}(\Vec{k}_{rel},\Vec{K}_{CM})=n^{NN}({k}_{rel},{K}_{CM},\Theta)\,=\frac{1}{(2\pi)^6}\int
d\Vec{r}\,d\Vec{R}\,d\Vec{r}^\prime\,d\Vec{R}^\prime\,
e^{i\,\Vec{K}_{CM}\cdot\left(\Vec{R}-\Vec{R}^\prime\right)}\,
e^{i\,\Vec{k}_{rel}\cdot\left(\Vec{r}-\Vec{r}^\prime\right)}\,
\rho^{(2)}_{NN}(\Vec{r},\Vec{R};\Vec{r}^\prime,\Vec{R}^\prime),
\label{2brelCM} \eeqy
\end{widetext}
where $|\Vec{k}_{rel}| \equiv k_{rel}$, $|\Vec{K}_{CM}| \equiv
K_{CM}$  and $\Theta$ is the angle between $\Vec{k}_{rel}$ and
${\Vec K}_{CM}$. In what follows the momentum distributions are
normalized to unity. Given the formula above, the momentum
distribution integrated over the CM coordinate,
$n^{NN}_{rel}(k_{rel})$, and the one integrated over the relative
momentum, $n^{NN}_{CM}(K_{CM})$, can be obtained, but a more
important property, considered in this Letter, is the dependence
of the
\begin{figure}[!htp]
\centerline{
  \includegraphics[width=9.3cm]{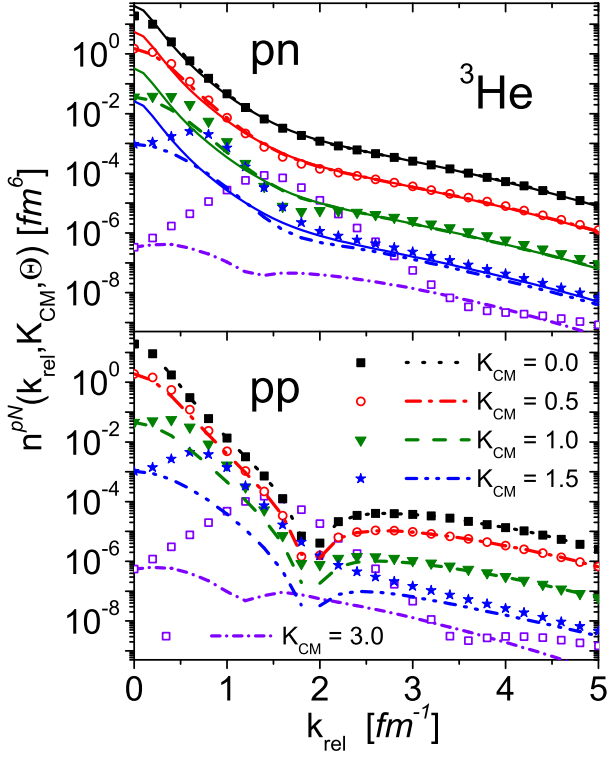}}
  \vskip -0.6cm
  \caption{ The  two-body momentum
  distributions   of   $pn$    and $pp$  pairs in $^3He$ normalized to unity, {\it vs.} the  relative momentum
  $k_{rel}$,  for fixed values of the
   CM momentum $K_{CM}$ and two orientations of them:
   ${\bf k}_{rel}
    ||{\bf K}_{CM}$ ({\it broken curves}) and ${\bf k}_{rel} \perp {\bf
    K}_{CM}$ ({\it symbols}). The continuous curves for the $pn$ pair represents the deuteron momentum distribution
    rescaled by the CM momentum distribution
    $n_{CM}^{pn}(K_{CM})= \int n^{pn}(\Vec{k}_{rel},\Vec{K}_{CM})\, d\Vec{k}_{rel}$ (see text and Fig. \ref{Fig4}).
  $^3He$  wave function from Ref. \cite{Kievsky:1992um} and
  $AV18$ interaction \cite{Wiringa:1994wb}.}
  \label{Fig1}
\end{figure}
two-body momentum distribution  upon the relative momentum
$k_{rel}$ for fixed values of the CM momentum $K_{CM}$  and the
angle $\Theta$.
\begin{figure}[!htp]
\centerline{
  \includegraphics[width=9.3cm]{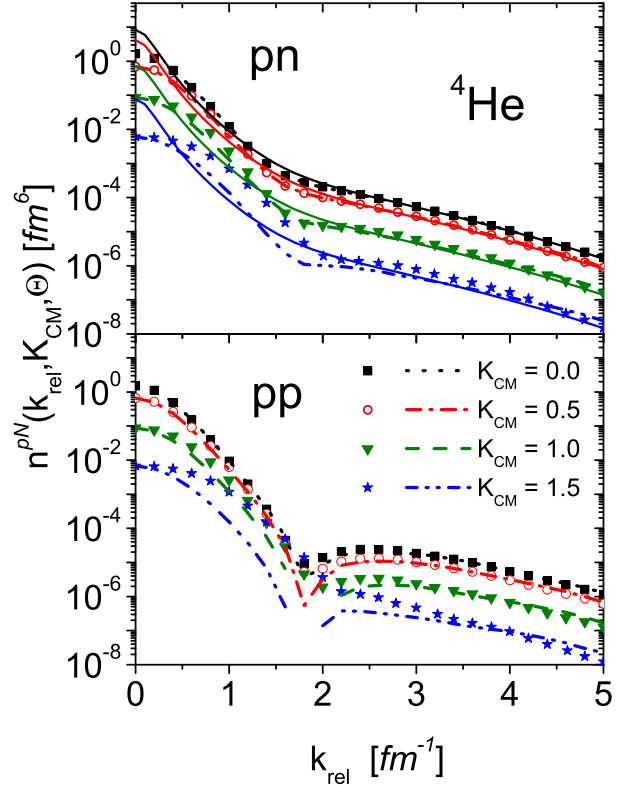}}
  \vskip -0.6cm
  \caption{The same as in Fig. \ref{Fig1} but for $^4He$. Correlated variational  wave function  from
  \cite{akaishi} and
    $AV8'$ interaction \cite{Pudliner:1997ck}.}
  \label{Fig2}
\end{figure}
\begin{figure}[!htp]
  \centerline{\includegraphics[width=8.5cm]{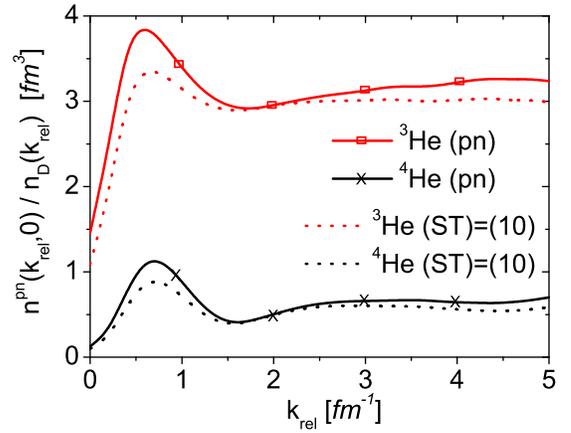}}
  \vskip -0.6cm
  \caption{The ratio of the $pn$  momentum distributions
  at at $K_{CM}=0$
   to the deuteron momentum distributions $n_D(k_{rel})$. The contribution from the
    spin-isospin deuteron states $S=1$, $T=0$ is also shown.}
    \label{Fig3}
\end{figure}
\begin{figure}[!htp]
  \centerline{\includegraphics[width=9.0cm]{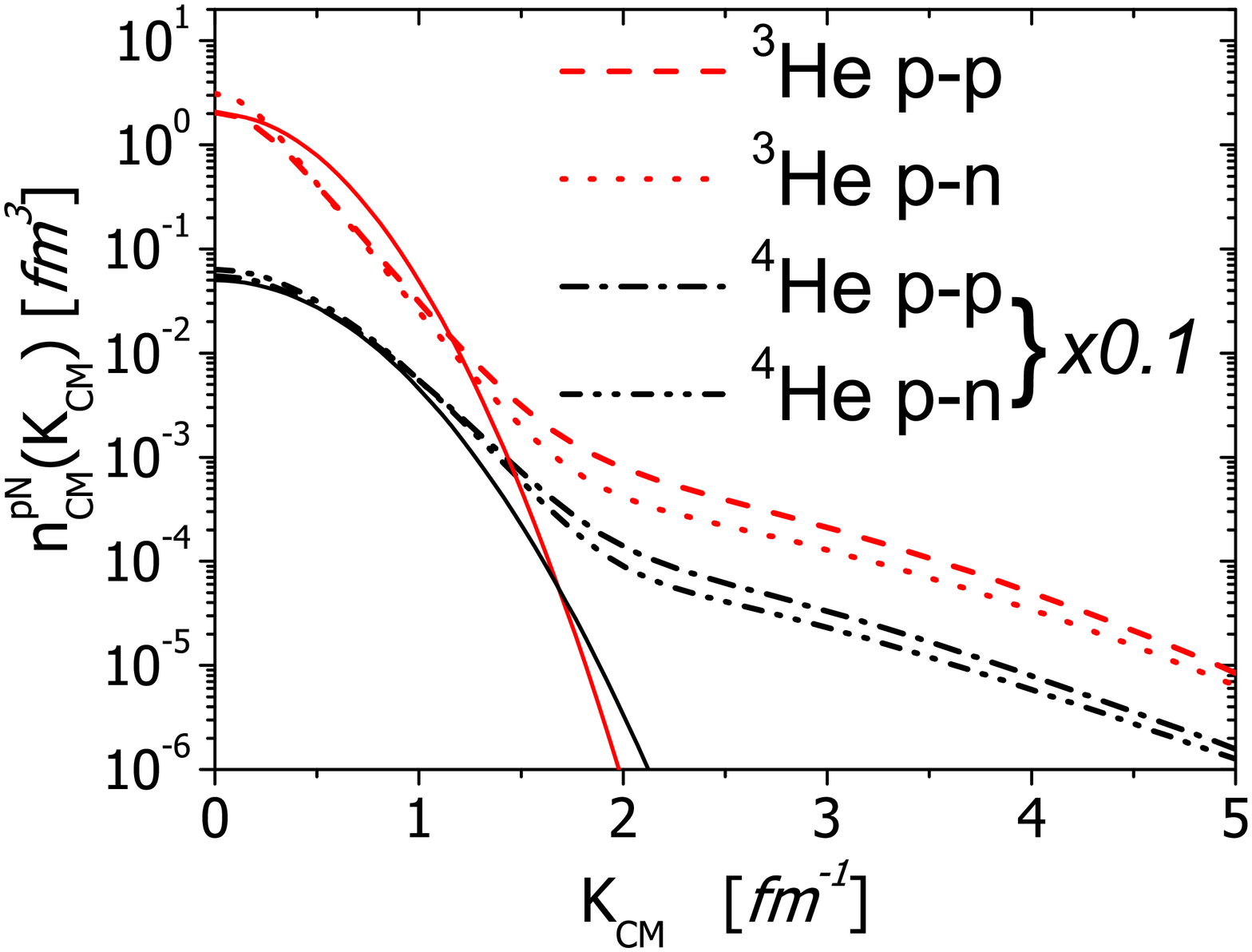}}
  \vskip -0.6cm
  \caption{The CM momentum distribution $n_{CM}^{pN}(K_{CM})= \int n^{pN}(\Vec{k}_{rel},\Vec{K}_{CM})\, d\,^3 k_{rel}$
    for $pp$ and $pn$ pairs in $^3He$ and $^4He$. The  solid lines correspond to the model
    of Ref. \cite{CiofidegliAtti:1995qe} aimed at describing the low momentum  ($K_{CM} \lesssim 1\,fm^{-1}$)
    part of $n_{CM}^{pN}(K_{CM})$.}
  \label{Fig4}
\end{figure}
In our calculations we  used,  for $^3He$, the nuclear wave
function obtained within the  approach from Ref.
\cite{Kievsky:1992um} and corresponding to the $AV18$ interaction
\cite{Wiringa:1994wb} and, for $^4He$   the  wave functions of
Ref. \cite{akaishi} corresponding to
 the $AV8'$ interaction \cite{Pudliner:1997ck}. Before discussing our results, let us stress
that the independence of the two-nucleon momentum distributions
upon the angle $\Theta$,  is evidence of the factorization of the
distributions in the variables $k_{rel}$ and $K_{CM}$,  i.e.
$n^{NN}({k}_{rel}, {K}_{CM},\Theta) \simeq
n_{rel}^{NN}({k}_{rel})n_{CM}^{NN}({K}_{CM})$
\cite{CiofidegliAtti:2010xv,Baldo:1900zz}. The $pn$ and $pp$
relative momentum distributions,  plotted {\it vs.} $k_{rel}$ in
correspondence of several values of $K_{CM}$ and two angular
configurations are shown in Figs.~\ref{Fig1} and~\ref{Fig2}; the
ratio $R^{pn}= n^{pn}(k_{rel},0)/n_D(k_{rel})$  for back-to-back
nucleons ($n^{pn}(k_{rel},0) \equiv n^{pn}(k_{rel},K_{CM}=0)$) is
presented in Fig.~\ref{Fig3}, whereas the $pN$ CM momentum
distributions $n_{CM}^{pN}(K_{CM})=\int
\,n^{pN}(\Vec{k}_{rel},\Vec{K}_{CM})d\Vec{ k}_{rel}$ are given  in
Fig.~\ref{Fig4}; finally, in Fig.~\ref{Fig5} the ratio $R_{pp/pn}$
of the correlated $pp$ to $pn$ pairs, extracted from the
$^3He(e,e^\prime pp)n$ process~\cite{Weinstein08}, is shown.

The main features of our results can be summarized as follows:
\begin{figure}[!htp]
  \centerline{\includegraphics[width=9.0cm]{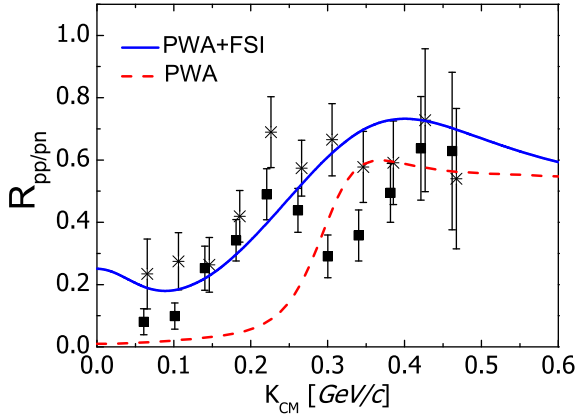}}
  \vskip -0.6cm
  \caption{The ratio of  the spectator correlated $pp$
and $pn$ nucleon pairs  extracted from the $^3He(e,e^{\prime}pp)n$
reaction and  integrated over the pair relative momentum  in the
range $1.5 <k_{rel} < 3.0\,\, fm^{-1}$ and the  angle $\Theta$
between ${\bf K}_{CM}$ and ${\bf k}_{rel}$ \cite{Weinstein08}. The
dashed curve represents the ratio of our calculated momentum
distributions and the full curve also includes the effect of the
$pp$ and $pn$ rescattering in the final state.}
  \label{Fig5}
\end{figure}
(i) at $K_{CM}=0$   the results of Ref. \cite{Schiavilla:2006xx}
are reproduced, namely  at  small values of $k_{rel}$ the $pn$ and
$pp$ momentum distributions do not appreciably differ, with their
ratio  being closer to the ratio of the $pn$ to $pp$ pairs,
whereas at $1.0\lesssim k_{rel} \lesssim 4.0\,\,fm^{-1}$   the
dominant role of  tensor correlations makes the $pn$
distributions  much larger than   $pp$ distribution, with the node
exhibited by the latter filled up by the $D$ wave in the $pn$
two-body density; (ii) $n^{NN}({k}_{rel},K_{CM}, \Theta)$, plotted
{\it vs.} $k_{rel}$, decreases with increasing values of $K_{CM}$;
  (iii)  starting from a given value of
$k_{rel}$, which for $K_{CM}=0$ is $k_{rel}\simeq 1.5\,\,
fm^{-1}$, and increases  with increasing
 $K_{CM}$, the $pn$ distribution
 changes its slope and becomes close to the
deuteron distribution; (iv) in the  region (${k}_{rel}\gtrsim 2
\,fm^{-1},{K}_{CM}\lesssim 1\,fm^{-1}$), $n^{NN}$ becomes
$\Theta$-independent, \footnote{Such an independence has been
checked in a wide range of angles}, which means that
$n^{NN}({k}_{rel},{K}_{CM},\Theta)\simeq n_{rel}^{NN}({k}_{rel})
n_{CM}^{NN}(K_{CM})$; for  $pn$ pairs, one
$n^{pn}({k}_{rel},{K}_{CM},\Theta) \simeq
n_D(k_{rel})n_{CM}^{pn}({K}_{CM})$, where $n_D(k_{rel})$ is the
deuteron momentum distribution and the only A-dependence is given
by  $n_{CM}^{pn/A}({K}_{CM})$; the factorized  form for $pn$ pairs
 describes the 2N SRC configuration, when the relative
momentum of the pair is much larger  than the CM momentum; (v) at
high values of the CM momentum, of the same order   of the (large)
relative momentum, more than two particles can be locally
correlated, with a resulting strong dependence upon the angle and
the breaking down of factorization, as clearly shown by Fig.
\ref{Fig1} for $K_{CM}= 3\,\,fm^{-1}$.  According to our
preliminary results \cite{complex}, all of the above remarks
appear to hold also for complex nuclei. Let us now discuss in
detail the factorized form of the momentum distributions for  $pn$
pairs. To this end we will consider the ratio $R^{pn}=
n^{pn}(k_{rel},0)/n_D(k_{rel})$   and its isospin dependence,
presented in Fig. \ref{Fig3},  and the CM momentum distribution
$n_{CM}^{pN}(K_{CM})$, presented in Fig. \ref{Fig4}. These two
Figures tell us, first of all, that the constant value exhibited
by the $S=1$, $T=0$ ratio at  $k_{rel}\gtrsim 1.5 \,\,fm^{-1}$, is
unquestionable evidence that  in this region the  dependence upon
$k_{rel}$ of the two body momentum distribution
$n^{pn}(k_{rel},0)$ is the same as  the deuteron one; secondly,
they also tell us that the difference between the ratios for
$^3He$ and $^4He$ in the region $k_{rel} \gtrsim 1.5 \,\,fm^{-1}$
equals exactly the difference between the values of the CM
momentum distributions at $K_{CM}=0$, shown in Fig. \ref{Fig4}; as
a consequence,  if we divide the dotted lines by the corresponding
values of $n_{CM}^{pn}(0)$, we obtain $1$ for both nuclei.
Concerning the different behavior of $n_{CM}^{pn}({K}_{CM})$ for
$^3He$ and $^4He$ at $K_{CM}\lesssim 1.5 \,\,fm^{-1}$, this is due
to the different binding associated with the CM motion: in $^3He$
the third uncorrelated particle is weakly bound, with a long
asymptotic tail, resulting in a sharp peak at $K_{CM}=0$;  thus
the more rapid fall off of the  CM momentum distributions of
$^3He$ leads,  with respect to the $^4He$ case, to the wider
separation of the curves corresponding to various values of
$K_{CM}$  presented in Fig. \ref{Fig1}. In $^4He$, the overall
average density can already be described by a mean field approach,
so that the realistic calculation leads, as shown in Fig.
\ref{Fig4}, to a result which is practically the same as the one
obtained in Ref. \cite{CiofidegliAtti:1995qe} within a model based
upon  the mean value of the kinetic energy in a shell model
picture. As for the experimental ratio  presented in Fig.
\ref{Fig5}, it should be pointed out that whereas this quantity
represents a nice confirmation of the dominance of tensor
correlations, it cannot provide information about the increase or
decrease of the two-body momentum distributions with the increase
of the CM momentum, since the $pp$ and $pn$ distributions may both
increase or decrease at the same time, leaving the ratio almost
unchanged. A discriminating quantity would be the ratio of $pn$
(or $pp$) pairs in correspondence of two values of the CM
momentum. As a matter of fact, it can be seen from Fig. \ref{Fig1}
that such a ratio at  e.g., $K_{CM}=0$ and
$K_{CM}=1.5\,\,fm^{-1}$, is predicted to be  a large positive
number.

To sum up, a clear physical picture of the motion of a pair of
nucleons embedded in the nuclear medium arises from our
calculations. In the region $2 \lesssim {k}_{rel}\lesssim 5
\,\,fm^{-1}$, ${K}_{CM}\lesssim 1\,fm^{-1}$, the motion of $NN$
pairs  is governed by 2N SRC, characterized by a decoupling of the
 CM and relative
 motions; for a $pn$ pair,
 the latter is described by the deuteron momentum distribution and the former is  governed by the average mean field motion. Some
 aspects of this picture have already been experimentally confirmed \cite{Subedi:2008zz}, whereas some others, e.g. the CM dependence of two
  nucleon momentum distributions,
 need  proper experimental investigations. This  picture  of a locally correlated pair,
 with the relative motion being practically
  A-independent,
 with the A-dependence given only by the CM motion, would be of great usefulness in
 various fields
   where SRC correlations have been recently shown to play an important role, such
   as
    high energy
  hadron-nucleus \cite{CiofidegliAtti:2011fh} and nucleus-nucleus
  scattering \cite{Alvioli:2010yk}, deep inelastic scattering \cite{Piasetzky:2011zz},
  the equation of state of nuclear \cite{Frick:2004th} and neutron  \cite{Frankfurt:2008zv}
  matters.

\section{acknowledgment}

HM and LPK thanks INFN, Sezione di Perugia, for kind hospitality.
The Work of MA is supported by the project HadronPhysics2 of the
European Commission, grant number 227431. Calculations were
performed at CASPUR,  thanks to the Standard HPC Grants 2010
SRCNuc and 2011 SRCNuc2.

\end{document}